# Learning the right channel in multimodal imaging: automated experiment in Piezoresponse Force Microscopy

Yongtao Liu,[1] Rama K. Vasudevan,[1] Kyle P. Kelley,[1] Hiroshi Funakubo,[2] Maxim Ziatdinov,[1,3,a] and Sergei V. Kalinin[4,b]

[1] Center for Nanophase Materials Sciences, Oak Ridge National Laboratory, Oak Ridge, TN 37923, USA

[2] Department of Material Science and Engineering, Tokyo Institute of Technology, Yokohama 226-8502, Japan

[3] Computational Sciences and Engineering Division, Oak Ridge National Laboratory, Oak Ridge, TN 37923, USA

[4] Department of Materials Science and Engineering, University of Tennessee, Knoxville, TN, 37996, USA

We report the development and experimental implementation of the automated experiment workflows for the identification of the best predictive channel for a phenomenon of interest in spectroscopic measurements. The approach is based on the combination of ensembled deep kernel learning for probabilistic predictions and a basic reinforcement learning policy for channel selection. It allows the identification of which of the available observational channels, sampled sequentially, are most predictive of selected behaviors, and hence have the strongest correlations. We implement this approach for multimodal imaging in Piezoresponse Force Microscopy (PFM), with the behaviors of interest manifesting in piezoresponse spectroscopy. We illustrate the best predictive channel for polarization-voltage hysteresis loop and frequency-voltage hysteresis loop areas is amplitude in the model samples. The same workflow and code are universal and applicable for any multimodal imaging and local characterization methods.

[a] ziatdinovma@ornl.gov
[b] sergei2@utk.edu

Multimodal imaging methods underpin multiple areas of fundamental and applied sciences. Conventional intermittent contact mode Atomic Force Microscopy yields topographic, phase, and error signals that highlight different aspects of surface structure.[1-3] In combination with detection modes such as Electrostatic,[4-6] Magnetic,[7-9] and Kelvin Probe Force Microscopy,[10-14] these technique offers multiple information channels containing information on dissimilar aspects of materials functionality. In optical imaging in biology, specific dies are used to highlight different elements of cell structure and are visualized with different color filters or spectral range in hyperspectral methods. In energy-dispersive electron microscopy and electron energy loss spectroscopy (EELS),[15, 16] different energy ranges highlight concentrations of individual elements.[17]

In many cases, imaging is used to define objects of interest for more detailed studies.[18-21] In scanning probe microscopy (SPM), the structural or functional images can be used to select locations for force-distance or current-voltage measurements,[22] or locations for local sampling for chemical studies. In optical and scanning electron microscopy, the imaging data can be used to select locations for e.g., nanoindentation.[23] In mass-spectrometry, the sampling points are often selected based on the optical or SPM imaging.[24, 25] This paradigm of imaging followed by selection of specific location(s) for detailed studies is common across physical, chemical, and biological imaging. Currently, these studies are often performed as guided by human operator intuition, via a classical point and click approach. However, in this case the process is slow and heavily biased by operator experience and expectations. An alternative approach is that of dense grid-based measurements, such as force-volume,[26] piezoresponse spectroscopy, piezoresponse nonlinearity measurements in SPM,[27-31] hyperspectral electron energy loss spectroscopy (EELS) measurements in scanning transmission electron microscopy (STEM),[32] photoluminescence lifetime measurement in optical microscopy,[19] or electron diffraction measurement in electron microscopy.[29] However, the grid measurements tend to be time consuming and are often limited or impossible for circumstances where the probe or the sample degrade rapidly with measurements.

An alternative in multimodal imaging is thus naturally of interest, enabling a spectroscopy workflow within an automated experiment framework. In this framework, the locations for spectroscopic studies are selected based on the features of interest in multimodal image. Here, the direct problem—performing measurements at a known location of interest—can be engendered via (by now) standard computer vision algorithms. For example, we can choose the specific objects such as domain walls or molecules, to identify locations for detailed spectroscopic measurements.[18, 21, 33-35]

However, the inverse problem—discovering the features of interest in the right channel, e.g., topography, or piezoresponse, or conductivity image channel, that are best predictive of behaviors of interest—is poorly amenable to human operation. For example, we aim to discover which microstructural element has the best predictive capacity for the functional property encoded in polarization hysteresis loop or resonance frequency hysteresis loop such as maximal loop area, imprint bias, or more complex functionals of the loop shape. For unimodal imaging, this approach have recently been demonstrated for STEM-EELS, 4D STEM, and band excitation piezoresponse spectroscopy (BEPS).[27, 32, 36] In these studies, we have discovered which features in image space are most predictive of the specific functionalities determined via spectral measurements, for example localization of the hysteresis loops with the maximal area at specific domain walls or emergence of low energy plasmons at the edges of 2D material flakes.

Here, we develop a framework for the automated discovery of the best predictive channel in multimodal imaging for the behavior of interest within a spectroscopic data set. Traditionally, such analysis is based on physical intuition using *a priori* expected physical relationships. However, this approach often leads to significant operator biases and precludes the discovery of the phenomena of interest. Here, we develop the experimental framework towards the discovery of the channel that offers best predictability for the behavior of interest in multimodal imaging. We have chosen to illustrate these using Piezoresponse Force Microscopy (PFM) as the method that allows multichannel imaging and extensive set of spectroscopies.[37] However, this approach is universal and applies to other forms of multimodal imaging.

As a model system, we explored three thin film samples: lead titanate (PTO),[38] lead zirconate titanate (PZT), and bismuth ferrite (BFO), these films are grown on $SrRuO_3$ layers. Band excitation piezoresponse force microscopy (BEPFM) measurements were performed on three model thin film materials to investigate their domain structure. These results are shown in Figure 1. The PTO thin film indicates both 180° ferroelectric domain structures—dark domain and bright domain in phase image (Figure 1b), and non-180° ferroelastic domain structures—dark and bright stripe domains in amplitude image (Figure 1a). The ferroelastic domains exhibit different strain and elastic properties due to the variation in crystallographic orientation, resulting in visible domain contrast in resonance frequency image (Figure 1c). Topography image (Figure 1d) also illustrates the ferroelastic domain features. In contrast, the PZT thin film only exhibits non-180º ferroelastic domain structures, displaying in BEPFM amplitude (Figure 1e), phase (Figure 1f), resonance frequency (Figure 1g), and topography (Figure 1h) images. The BFO majorly shows 180º ferroelectric domain structure (Figure 1i-j), where the domain wall contrast is also visible in resonance frequency image (Figure 1k). Notably, a few ferroelastic domains with weak contrast also show in amplitude image (Figure 1i).

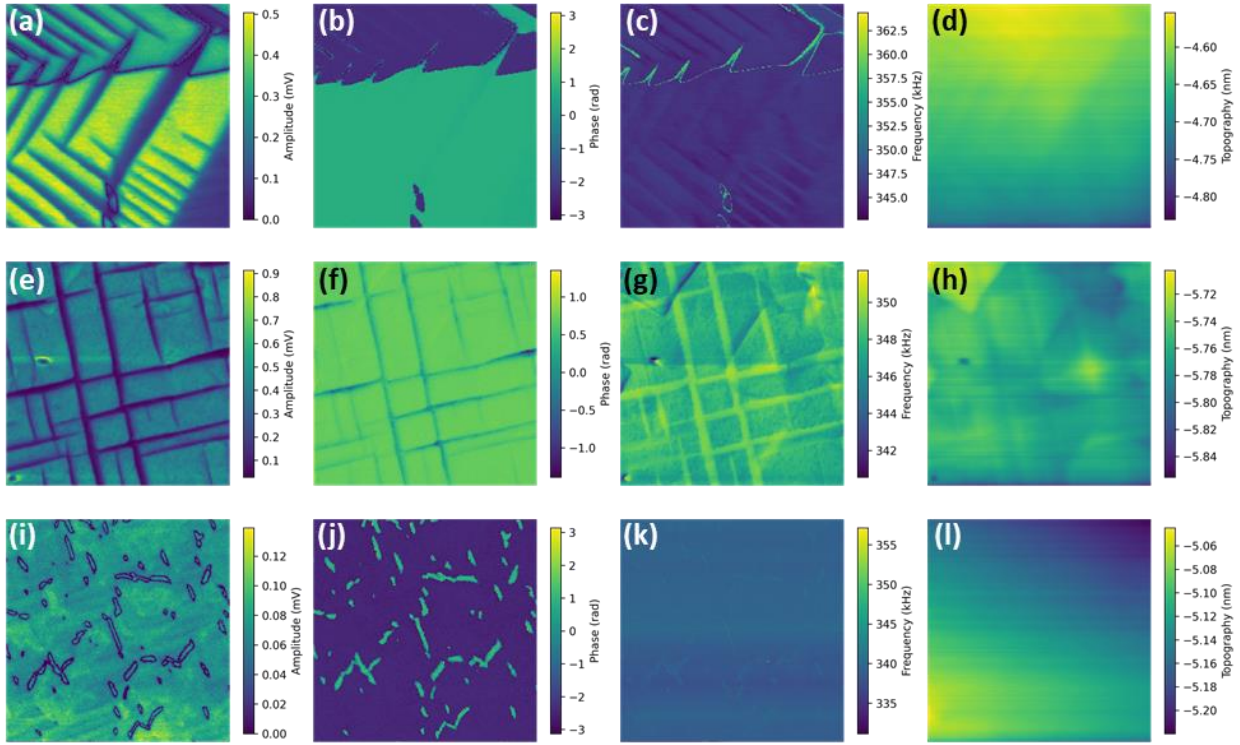

**Figure 1.** Band excitation piezoresponse force microscopy (BEPFM) image results of three model samples. (a-d), BEPFM amplitude, phase, resonance frequency, and topography images of PTO sample, respectively. (e-h), BEPFM amplitude, phase, resonance frequency, and topography images of PZT sample, respectively. (i-l), BEPFM amplitude, phase, resonance frequency, and topography images of BFO sample, respectively.

Next, we perform a multiple-channel deep kernel learning (DKL) measurement utilizing the ensembles of DKL models and basic reinforcement learning policy. Earlier, we showed how combining the structured Gaussian process with the epsilon-greedy policy allows one to learn a correct model of the system's behavior and use it to drive the exploration of the configuration space.[39, 40] However, that approach is limited to low-dimensional spaces and is not suitable for the structure-property relationship problems in the multimodal imaging. Here we use DKL[41] that is a hybrid of a neural network and a Gaussian process to circumvent the dimensionality problem. As the fully Bayesian implementation of DKL is computationally too slow for real-time feedback and control, we approximated it with the ensembles of DKL models.[42] In this setup, each neural network in the ensemble is initialized independently resulting in different embeddings connected to separate Gaussian processes and the final prediction for each channel is an ensemble average.

The process of channel learning with ensemble-DKL is shown in Figure 2a-2b. The BEPFM images including amplitude, phase, frequency, and topography are used as four possible input channels. Each image is featurized by splitting it into patches that are used as inputs. The behavior of interest is encoded in polarization or resonance frequency hysteresis loops for each patch as a scalar target. Here we use the hysteresis loop area, but any functional of the spectroscopic signal can be selected. At the beginning of the channel learning experiment, a small, custom-defined number of warm-up steps is taken, at which a separate ensemble of DKL models is trained for each channel. In this process, the channel that produces the lowest mean predictive uncertainty on the unmeasured points is given a positive reward. This rewarded model is also used to derive the next measurement point corresponding to the largest uncertainty value in the prediction. After the warm-up steps, an epsilon-greedy policy[43] is used to sample a single channel at each exploration step and derive the next measurement point.

We implement this ensemble-DKL workflow on an Oxford Instrument Asylum Cypher microscope. As shown in Figure 2c, to accelerate the DKL training and prediction we send the real-time measurement data to an Nvidia DGX-2 GPU server for analysis. Specifically, the custom DKL code written in JAX[44] is run on a docker container residing on the GPU server. Via a combination of port forwarding and socket programming, data is sent directly from the instrument computer to the DGX-2 device without file I/O, and then processed within the container, taking advantage of the high processing capabilities on the server. For the data transfer, we utilize the mlsocket package,[45] which is a wrapper around the low-level python socket interface and enables sending and receiving of numpy arrays. The server houses 16 Nvidia Tesla V-100 GPUs each with 32GB of memory, enabling the different ensemble models to run in parallel. Practically, we select between multi-GPU "parallel" and single GPU "vectorized" approach for the ensemble-DKL training based on the size of image patch and complexity/depth of the neural network. For the image patch size of 20x20, the 3-layer fully-connected neural network, and 20 ensemble models, each iteration takes ~30s when utilizing a single GPU, whereas for a comparison, the same iteration takes ~300s on the CPU. As such, the connection to edge computing is critical for efficiency and viability of the proposed workflow.

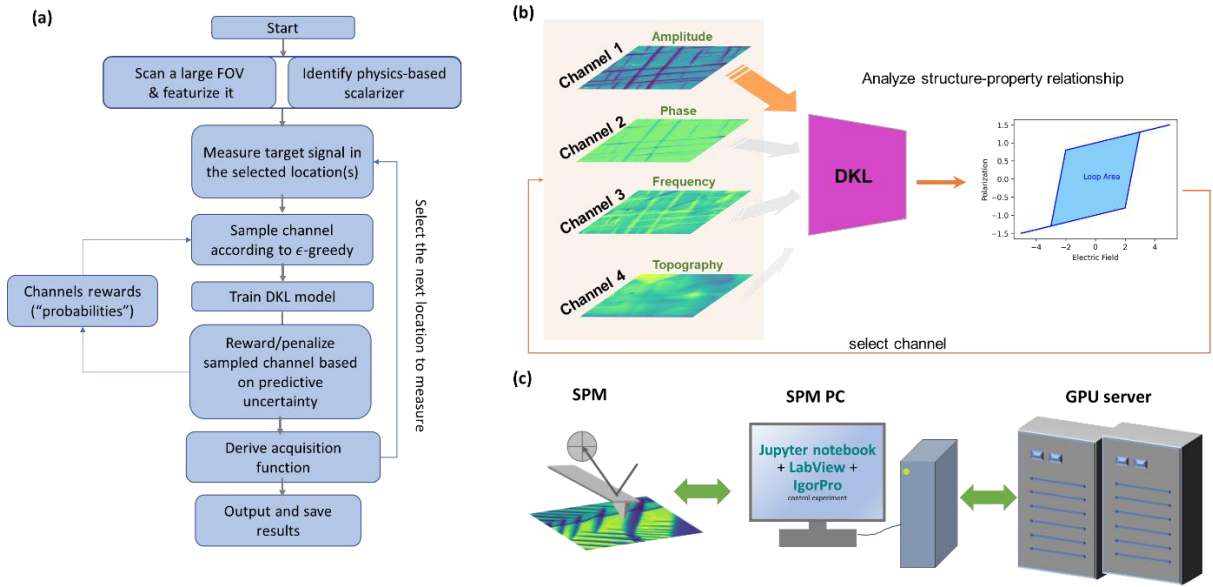

**Figure 2.** BEPFM experimental process driven by ensemble DKL. (a) Ensembel DKL workflow. (b) BEPFM image channels are used to predict a functional property, e.g., polarization loop area or resonance frequency loop area. The image data has four channels: amplitude (Channel 1), phase (Channel 2), resonance frequency (Channel 3), and topography (Channel 4). The goal is to identify the best channel for predicting the functional property. (c) a schematic showing hardware connected in the workflow, including a Cypher SPM, a PC, and a GPU server.

Here, we performed two sets of measurements—the polarization-voltage loop area and frequency-voltage loop area are used as target property descriptors, which measure the energy loss during switching and voltage-induced irreversible dynamics, respectively—on three model samples. First, a small number of randomly sampled points are measured as seed points for training. In these measurements, we start with 0.25% of the total measurement points as the seed data for DKL training, then perform 20 warm-up steps and 200 exploration steps. In the warm-up steps, each channel is trained in parallel and the one with the lowest mean uncertainty is used to derive the next measurement point. After the warm-up, a single channel is sampled at each step according to the epsilon-greedy policy with epsilon decreased uniformly ("annealed") from 0.4 to 0.1 during the 200 exploration steps.

Shown in Figure 3 are the evolution of channel reward, mean predictive uncertainty, and channel selection during the ensemble-DKL driven measurement for three samples. For the PTO sample, when the target property is polarization-voltage loop area (Figure 3a), amplitude channel shows the highest reward and the phase channel is the second-best. Although the resonance frequency shows a very low reward (Figure 3a), the evolution of uncertainty (Figure 3b) indicates that the predictive uncertainty based on resonance frequency channel gradually decreases during the experiment, which implicates that the elastic variation displayed in frequency image has a role in polarization dynamics. However, the topography channel shows both low reward (Figure 3a) and no decrease of prediction uncertainty (Figure 3b). When the frequency-voltage loop area is used as the target property, we observe an increase of reward to the resonance frequency and phase channels at the end of experiment (Figure 3c), accompanied with larger decrease rate of predictive uncertainty from the resonance frequency and phase channels (Figure 3d). The behavior of resonance frequency channel is due to the directly correlated property from loops and image data. The behavior of phase channel can be understood as the electrostatic effect on the detected cantilever resonance frequency, where the up and down polarized domains (shown as dark and bright contrast in phase image) may associate with different surface charge states that induce different electrostatic effect.

The PZT results (Figure 3e-h) is very similar to those of PTO. We ascribe this similarity to the fact that most variability of the phenomena on epitaxial film surfaces are related to ferroelastic domain structure. However, note that predictive uncertainty from topography channel (Figure 3f and 3h) slightly decreases during experiment in the PZT sample.

For the BFO results (Figure 3i-l), when the polarization-voltage loop area is used as target property, the reward to amplitude channel (Figure 3i) quickly stabilized around 0.5-0.6 after ~50 exploration steps, while other channel rewards drop quickly. Interestingly, the predictive uncertainties of four channels are distinct (Figure 3j)—the uncertainty corresponding to amplitude channel keeps decreasing, the phase channel uncertainty is also very low but shows a slight increase in the middle of the measurement, and the uncertainties corresponding to frequency and topography channels are very high. When the resonance frequency-voltage loop area is used as target property, the evolution of channel reward and uncertainty (Figure 3k-l) is similar to that of polarization-voltage loop area as target property. This is most likely because both phenomena are ferroelectric domain related.

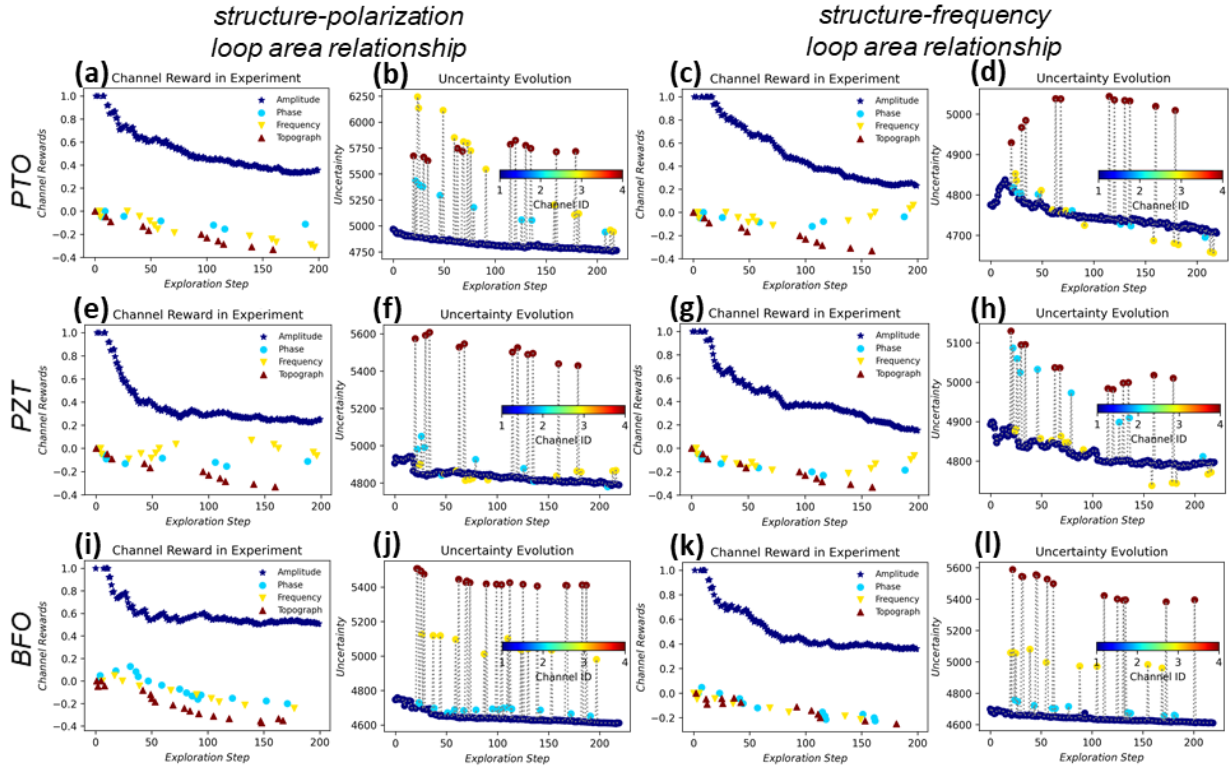

**Figure 3.** Experimental process of three model samples when the functional property is set as polarization loop area and resonance frequency loop area. (a-d), (e-h), (i-l), results for PTO, PZT, and BFO samples, respectively. The first and second columns show the evolution of channel reward and the mean predictive uncertainty as a function of experimental steps for the ensemble DKL when the functional property is a polarization loop area. In this case, ensemble DKL identifies the channel with best structure-polarization loop area relationship. The third and fourth columns show the evolution of channel reward and the mean predictive uncertainty as a function of experimental steps for the ensemble DKL when the functional property is resonance frequency loop area. In this case, the ensemble DKL identifies the channel with best structure-frequency loop area relationship. In these measurements, the ensemble DKL analysis starts with 0.25 % of total number of points available for measurements and use 20 warm-up states. During the warm-up states, all four channels are evaluated in parallel and the one with the lowest uncertainty is used for next evaluation point. After the warmup phase, we sample a single channel at each step based

on the epsilon-greedy policy with epsilon decreased uniformly from 0.4 to 0.1 during the 200 steps. In the uncertainty evolution plots, the color represents the selected channel at a specific step, which allows us to visualize the correlation of channel selection and uncertainty changes.

After the ensemble-DKL exploration measurement, we can use the ensemble-DKL model to predict the target property at unmeasured points. Notably, the prediction can be made from each channel. Shown in Figure 4 and Figure 5 are the prediction of polarization-voltage loop area and frequency-voltage loop area of three samples from each channel, respectively. For the PTO and PZT samples, predictions from topography (Figure 4d, 4h, and Figure 5d, 5h) display some features also showing up in the predictions from other channels, presumably because the ferroelastic domains also show in topography.

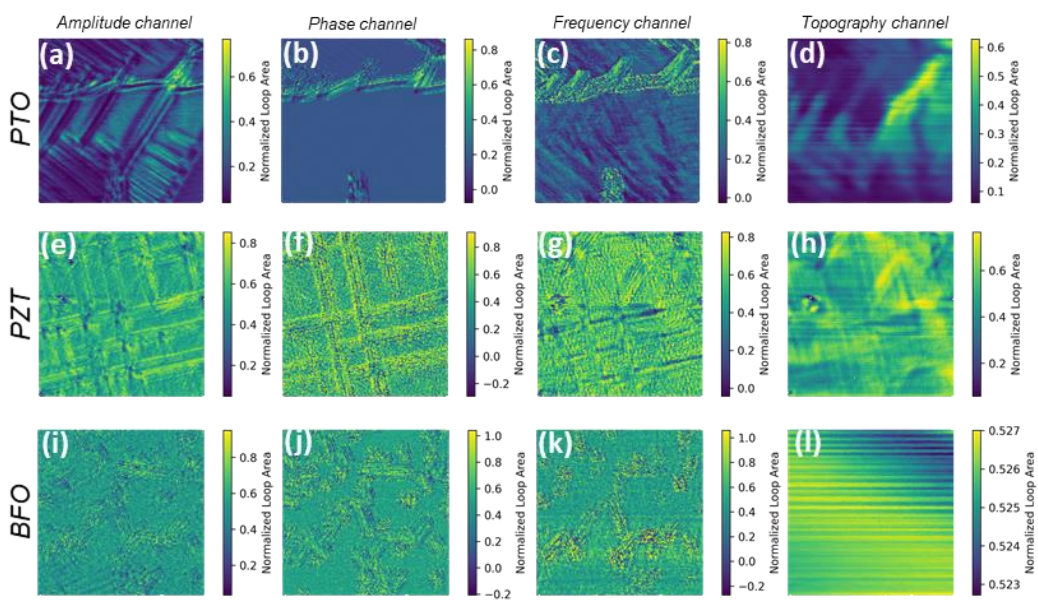

**Figure 4.** Ensemble DKL prediction of polarization loop area from each channel after experiment. (a-d), (e-h), (i-l), ensemble DKL predictions of PTO, PZT, and BFO, respectively. (a), (e), (i), DKL prediction from amplitude image; (b), (f), (j), DKL prediction of polarization loop area from phase image; (c), (g), (k), DKL prediction of polarization loop area from resonance frequency image; (d), (h), (l), DKL prediction of polarization loop area from topography image.

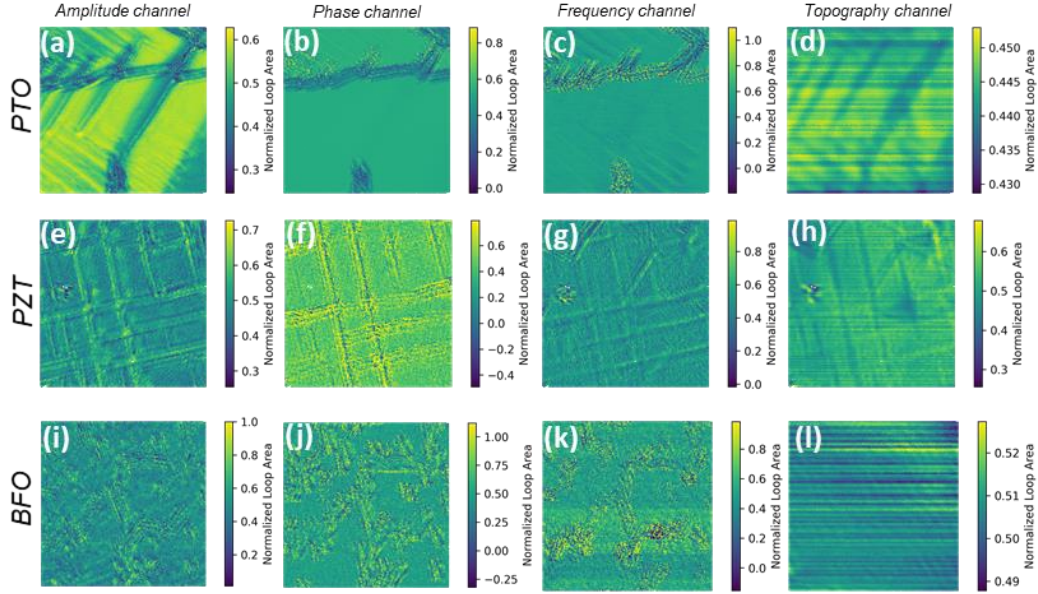

**Figure 5.** Ensemble DKL prediction of frequency loop area from each channel after experiment. (a-d), (e-h), (i-l), ensemble DKL predictions of PTO, PZT, and BFO, respectively. (a), (e), (i), DKL prediction of frequency loop area from amplitude image; (b), (f), (j), DKL prediction of frequency loop area from phase image; (c), (g), (k), DKL prediction of frequency loop area from resonance frequency image; (d), (h), (l), DKL prediction of frequency loop area from topography image.

The model selection during exploration steps is based on both the current channel reward (partially from warm-up steps) and the exploration/exploitation balance with epsilon-greedy policy. For the PTO and PZT results of using frequency-voltage loop area as target property, even if ensemble-DKL used the amplitude channel originally, the frequency channel reward starts increasing at the end of measurements (Figure 3c and Figure 3g) and the frequency channel uncertainty decreases faster than amplitude channel in some cases (Figure 3d and Figure 3h). Therefore, to investigate more details of the channel behaviors when using frequency-voltage loop area as target property, an additional experiment with enlarged exploration steps and different exploration/exploitation rate in epsilon-greedy policy was performed. In this measurement, we perform 20 warm-up steps and 480 exploration steps. In the exploration steps, epsilon in the epsilon-greedy policy decreased uniformly from 0.9 to 0.01 is used to sample a single channel at each step. Compared to previous measurements, here the epsilon is larger at the beginning of the measurement and smaller at the end of the measurement, corresponding to larger exploration rate at the beginning and smaller exploration rate at the end, respectively. Shown in Figure 6 are the results, in this measurement, other channels are used more frequently (Figure 6f). This is because of the higher exploration rate at the beginning of the measurement. In this case, we can observe more details of the evolution of other channels. We observe an obvious increase of reward to the phase channel (Figure 6e) and the fastest decrease of uncertainty from phase channel prediction (Figure 6f), probably because of the electrostatic effect as mentioned before.

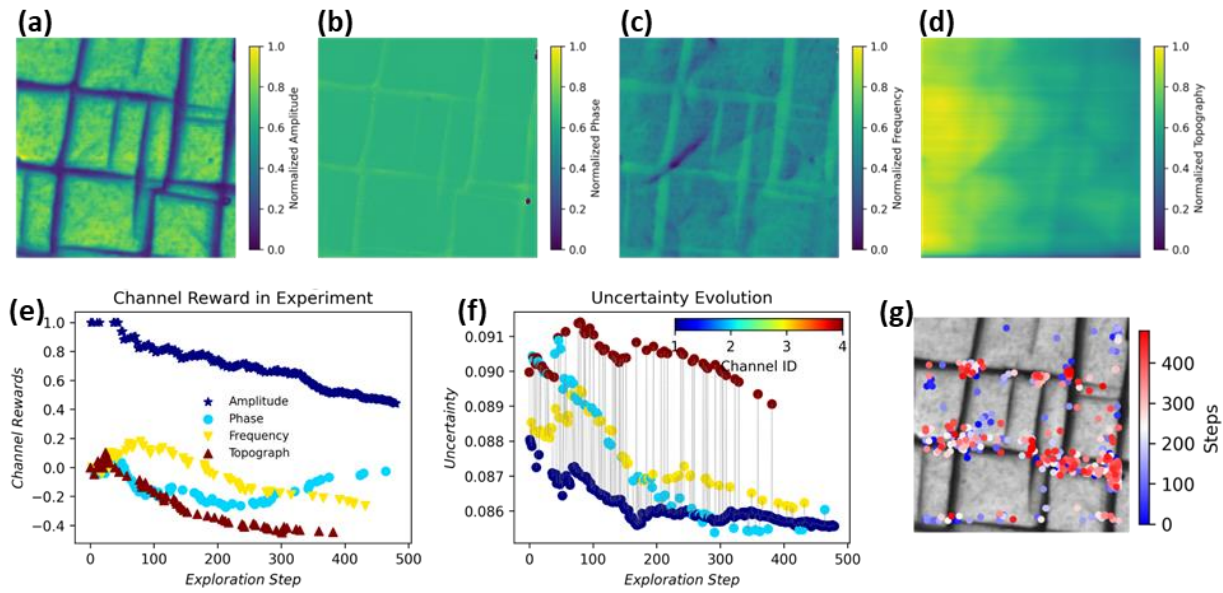

**Figure 6.** Experimental process exploring structure-resonance frequency loop area relationship in PZT sample with a different exploration and exploitation balance. (a-d), BEPFM amplitude, phase, resonance frequency, and topography images. (e) evolution of channel reward as a function of experiment steps. (f) the evolution of mean predictive uncertainty as a function of experimental reward. In this measurement, the ensemble DKL analysis starts with 0.4 % of the data and use 20 warm-up states. After warm-up, the channel at each step is sampled with the epsilon-greedy policy and the epsilon decreased uniformly from 0.9 to 0.01 during the 480 steps. Here, the different epsilon as compared with previous measurements lead to a higher exploration chance (due to larger epsilon) at the beginning and higher exploitation chance (due to smaller epsilon) at the end of the measurement. (g) the measurement locations determined by DKL showing on amplitude map. Note that the measurement points are concentrated at the a-x domain boundaries, at which the polarization tilting can give rise to enhanced responses.

To summarize, we have implemented an ensemble-DKL driven automated PFM for the identification of the channel with best predictive capacity, i.e., the channel for the most accurate reconstruction of target property encoded in spectroscopic data. This approach identifies the BEPFM image channel with the most predictive power for a target property of interest during measurement, which is also an indication of the strongest correlation between this BEPFM image channel and the target property.

Here, we implement this approach in BEPFM and piezoresponse spectroscopy measurement, and illustrate its application in exploring the structure-property relationships in three thin film materials with various ferroelectric and ferroelastic properties. To accelerate the ensemble-DKL training and prediction, we also develop an approach enabling real-time data transfer between microscope PC and GPU server, which allows GPU server to analyze the results from the on-the-fly microscope. This workflow and approach are universal and can be applied in other imaging and spectroscopic characterization methods, e.g. electron microscope, optical microscope, mass spectrometry imaging, as well.

**Acknowledgements**

This effort (implementation in SPM, measurement, data analysis) was primarily supported by the center for 3D Ferroelectric Microelectronics (3DFeM), an Energy Frontier Research Center funded by the U.S. Department of Energy (DOE), Office of Science, Basic Energy Sciences under Award Number DE-SC0021118. This research (ensemble-DKL) was supported by the Center for Nanophase Materials Sciences (CNMS), which is a US Department of Energy, Office of Science User Facility at Oak Ridge National Laboratory.

**Conflict of Interest Statement**

The authors declare no conflict of interest.

**Authors Contribution**

S.V.K. and M.Z. conceived the project. M.Z. realized the ensemble-DKL workflow. Y.L. implemented ensemble-DKL to SPM and obtained results. K.K. helped with the deployment. RKV assisted with connections to computational facilities, and analysis. H.F. provided the PTO and PZT sample. All authors contributed to discussions and the final manuscript.

**Data Availability Statement**

The method that support the findings of this study are available at https://github.com/yongtaoliu/Ensemble-DKL.

**Code Availability Statement**

The code related to this study are available at https://github.com/yongtaoliu/Ensemble-DKL.

# References

1. Gerber, C.; Lang, H. P., How the doors to the nanoworld were opened. *Nature Nanotechnology* **2006,** *1* (1), 3-5.
2. Garcia, R.; Perez, R., Dynamic atomic force microscopy methods. *Surface Science Reports* **2002,** *47* (6-8), 197-301.
3. Hong, J. W.; Park, S. I.; Khim, Z. G., Measurement of hardness, surface potential, and charge distribution with dynamic contact mode electrostatic force microscope. *Review of Scientific Instruments* **1999,** *70* (3), 1735-1739.
4. Coffey, D. C.; Ginger, D. S., Time-resolved electrostatic force microscopy of polymer solar cells. *Nature Materials* **2006,** *5* (9), 735-740.
5. Iwata, M.; Katsuraya, K.; Suzuki, I.; Maeda, M.; Yasuda, N.; Ishibashi, Y., Domain wall observation and dielectric anisotropy in PZN-PT by SPM. *Materials Science and Engineering B-Solid State Materials for Advanced Technology* **2005,** *120* (1-3), 88-90.
6. Ziegler, D.; Rychen, J.; Naujoks, N.; Stemmer, A., Compensating electrostatic forces by single-scan Kelvin probe force microscopy. *Nanotechnology* **2007,** *18* (22), 225505.
7. Martin, Y.; Wickramasinghe, H. K., MAGNETIC IMAGING BY FORCE MICROSCOPY WITH 1000-A RESOLUTION. *Applied Physics Letters* **1987,** *50* (20), 1455-1457.
8. Grutter, P.; Liu, Y.; LeBlanc, P.; Durig, U., Magnetic dissipation force microscopy. *Applied Physics Letters* **1997,** *71* (2), 279-281.
9. Popov, G.; Kalinin, S. V.; Alvarez, T.; Emge, T. J.; Greenblatt, M.; Bonnell, D. A., Micromagnetic and magnetoresistance studies of ferromagnetic $La_{0.83}Sr_{0.13}MnO_{2.98}$ crystals. *Phys. Rev. B* **2002,** *65* (6).
10. Nonnenmacher, M.; Oboyle, M. P.; Wickramasinghe, H. K., KELVIN PROBE FORCE MICROSCOPY. *Applied Physics Letters* **1991,** *58* (25), 2921-2923.
11. Tanimoto, M.; Vatel, O., Kelvin probe force microscopy for characterization of semiconductor devices and processes. *Journal of Vacuum Science & Technology B* **1996,** *14* (2), 1547-1551.
12. Baumgart, C.; Helm, M.; Schmidt, H., Quantitative dopant profiling in semiconductors: A Kelvin probe force microscopy model. *Phys. Rev. B* **2009,** *80* (8), 085305.
13. Sadewasser, S.; Jelinek, P.; Fang, C. K.; Custance, O.; Yamada, Y.; Sugimoto, Y.; Abe, M.; Morita, S., New Insights on Atomic-Resolution Frequency-Modulation Kelvin-Probe Force-Microscopy Imaging of Semiconductors. *Phys. Rev. Lett.* **2009,** *103* (26), 266103.
14. Melitz, W.; Shen, J.; Kummel, A. C.; Lee, S., Kelvin probe force microscopy and its application. *Surface Science Reports* **2011,** *66* (1), 1-27.
15. Bosman, M.; Watanabe, M.; Alexander, D. T. L.; Keast, V. J., Mapping chemical and bonding information using multivariate analysis of electron energy-loss spectrum images. *Ultramicroscopy* **2006,** *106* (11-12), 1024-1032.
16. Browning, N. D.; Buban, J. P.; Nellist, P. D.; Norton, D. P.; Chisholm, M. F.; Pennycook, S. J., The atomic origins of reduced critical currents at 001 tilt grain boundaries in $YBa_2Cu_3O_{7-\delta}$ thin films. *Physica C* **1998,** *294* (3-4), 183-193.
17. Kapetanakis, M. D.; Zhou, W.; Oxley, M. P.; Lee, J.; Prange, M. P.; Pennycook, S. J.; Idrobo, J. C.; Pantelides, S. T., Low-loss electron energy loss spectroscopy: An atomic-resolution complement to optical spectroscopies and application to graphene. *Phys. Rev. B* **2015,** *92* (12).
18. Liu, Y.; Fields, S. S.; Mimura, T.; Kelley, K. P.; Trolier-McKinstry, S.; Ihlefeld, J. F.; Kalinin, S. V., Exploring leakage in dielectric films via automated experiments in scanning probe microscopy. *Applied Physics Letters* **2022,** *120* (18), 182903.

19. Liu, Y.; Li, M.; Wang, M.; Collins, L.; Ievlev, A. V.; Jesse, S.; Xiao, K.; Hu, B.; Belianinov, A.; Ovchinnikova, O. S., Twin domains modulate light-matter interactions in metal halide perovskites. *APL Materials* **2020,** *8* (1), 011106.
20. Vasudevan, R. K.; Kelley, K. P.; Hinkle, J.; Funakubo, H.; Jesse, S.; Kalinin, S. V.; Ziatdinov, M., Autonomous experiments in scanning probe microscopy and spectroscopy: choosing where to explore polarization dynamics in ferroelectrics. *ACS nano* **2021,** *15* (7), 11253-11262.
21. Kelley, K. P.; Ren, Y.; Dasgupta, A.; Kavle, P.; Jesse, S.; Vasudevan, R. K.; Cao, Y.; Martin, L. W.; Kalinin, S. V., Probing Metastable Domain Dynamics via Automated Experimentation in Piezoresponse Force Microscopy. *ACS nano* **2021,** *15* (9), 15096-15103.
22. Cappella, B.; Dietler, G., Force-distance curves by atomic force microscopy. *Surface science reports* **1999,** *34* (1-3), 1-104.
23. Oliver, W. C.; Pharr, G. M., AN IMPROVED TECHNIQUE FOR DETERMINING HARDNESS AND ELASTIC-MODULUS USING LOAD AND DISPLACEMENT SENSING INDENTATION EXPERIMENTS. *Journal of Materials Research* **1992,** *7* (6), 1564-1583.
24. Liu, Y.; Borodinov, N.; Collins, L.; Ahmadi, M.; Kalinin, S. V.; Ovchinnikova, O. S.; Ievlev, A. V., Role of decomposition product ions in hysteretic behavior of metal halide perovskite. *ACS nano* **2021,** *15* (5), 9017-9026.
25. Liu, Y.; Ievlev, A. V.; Borodinov, N.; Lorenz, M.; Xiao, K.; Ahmadi, M.; Hu, B.; Kalinin, S. V.; Ovchinnikova, O. S., Direct observation of photoinduced ion migration in lead halide perovskites. *Advanced Functional Materials* **2021,** *31* (8), 2008777.
26. Gad, M.; Itoh, A.; Ikai, A., Mapping cell wall polysaccharides of living microbial cells using atomic force microscopy. *Cell Biology International* **1997,** *21* (11), 697-706.
27. Liu, Y.; Kelley, K. P.; Vasudevan, R. K.; Funakubo, H.; Ziatdinov, M. A.; Kalinin, S. V., Experimental discovery of structure–property relationships in ferroelectric materials via active learning. *Nature Machine Intelligence* **2022,** *4* (4), 341-350.
28. Liu, Y.; Vasudevan, R. K.; Kelley, K. K.; Kim, D.; Sharma, Y.; Ahmadi, M.; Kalinin, S. V.; Ziatdinov, M., Decoding the shift-invariant data: applications for band-excitation scanning probe microscopy. *Machine Learning: Science and Technology* **2021,** *2* (4), 045028.
29. Liu, Y.; Trimby, P.; Collins, L.; Ahmadi, M.; Winkelmann, A.; Proksch, R.; Ovchinnikova, O. S., Correlating crystallographic orientation and ferroic properties of twin domains in metal halide perovskites. *ACS nano* **2021,** *15* (4), 7139-7148.
30. Vasudevan, R. K.; Okatan, M. B.; Duan, C.; Ehara, Y.; Funakubo, H.; Kumar, A.; Jesse, S.; Chen, L. Q.; Kalinin, S. V.; Nagarajan, V., Nanoscale origins of nonlinear behavior in ferroic thin films. *Advanced Functional Materials* **2013,** *23* (1), 81-90.
31. K. Vasudevan, R.; Marincel, D.; Jesse, S.; Kim, Y.; Kumar, A.; V. Kalinin, S.; Trolier-McKinstry, S., Polarization dynamics in ferroelectric capacitors: local perspective on emergent collective behavior and memory effects. *Advanced Functional Materials* **2013,** *23* (20), 2490-2508.
32. Roccapriore, K. M.; Kalinin, S. V.; Ziatdinov, M., Physics discovery in nanoplasmonic systems via autonomous experiments in Scanning Transmission Electron Microscopy. *arXiv preprint arXiv:2108.03290* **2021**.
33. Kelley, K. P.; Ren, Y.; Morozovska, A. N.; Eliseev, E. A.; Ehara, Y.; Funakubo, H.; Giamarchi, T.; Balke, N.; Vasudevan, R. K.; Cao, Y., Dynamic manipulation in piezoresponse force microscopy: creating nonequilibrium phases with large electromechanical response. *ACS nano* **2020,** *14* (8), 10569-10577.
34. Sotres, J.; Boyd, H.; Gonzalez-Martinez, J. F., Enabling autonomous scanning probe microscopy imaging of single molecules with deep learning. *Nanoscale* **2021,** *13* (20), 9193-9203.
35. Huang, B.; Li, Z.; Li, J., An artificial intelligence atomic force microscope enabled by machine learning. *Nanoscale* **2018,** *10* (45), 21320-21326.

36. Roccapriore, K. M.; Dyck, O.; Oxley, M. P.; Ziatdinov, M.; Kalinin, S. V., Automated experiment in 4D-STEM: exploring emergent physics and structural behaviors. *ACS nano* **2022**.
37. Vasudevan, R. K.; Jesse, S.; Kim, Y.; Kumar, A.; Kalinin, S. V., Spectroscopic imaging in piezoresponse force microscopy: New opportunities for studying polarization dynamics in ferroelectrics and multiferroics. *Mrs Communications* **2012,** *2* (3), 61-73.
38. Morioka, H.; Yamada, T.; Tagantsev, A. K.; Ikariyama, R.; Nagasaki, T.; Kurosawa, T.; Funakubo, H., Suppressed polar distortion with enhanced Curie temperature in in-plane 90°-domain structure of a-axis oriented PbTiO3 Film. *Applied Physics Letters* **2015,** *106* (4), 042905.
39. Liu, Y.; Morozovska, A.; Eliseev, E.; Kelley, K. P.; Vasudevan, R.; Ziatdinov, M.; Kalinin, S. V., Hypothesis-Driven Automated Experiment in Scanning Probe Microscopy: Exploring the Domain Growth Laws in Ferroelectric Materials. *arXiv preprint arXiv:2202.01089* **2022**.
40. Ziatdinov, M. A.; Liu, Y.; Morozovska, A. N.; Eliseev, E. A.; Zhang, X.; Takeuchi, I.; Kalinin, S. V., Hypothesis learning in automated experiment: application to combinatorial materials libraries. *Advanced Materials* **2022**, 2201345.
41. Wilson, A. G.; Hu, Z.; Salakhutdinov, R.; Xing, E. P. In *Deep kernel learning*, Artificial intelligence and statistics, PMLR: 2016; pp 370-378.
42. Ziatdinov, M.; Liu, Y.; Kalinin, S. V., Active learning in open experimental environments: selecting the right information channel (s) based on predictability in deep kernel learning. *arXiv preprint arXiv:2203.10181* **2022**.
43. Zai, A.; Brown, B., *Deep reinforcement learning in action*. Manning Publications: 2020.
44. Bradbury, J.; Frostig, R.; Hawkins, P.; Johnson, M. J.; Leary, C.; Maclaurin, D.; Necula, G.; Paszke, A.; VanderPlas, J.; Wanderman-Milne, S., JAX: composable transformations of Python+ NumPy programs. *Version 0.2* **2018,** *5*, 14-24.
45. https://pypi.org/project/mlsocket/.